\documentstyle[aps,prl,twocolumn,epsf]{revtex}
\begin{document}
\draft
\preprint{}

\twocolumn[\hsize\textwidth\columnwidth\hsize\csname @twocolumnfalse\endcsname

\centerline {The Density Matrix Renormalization Group Studies of Metal-Halogen 
Chains}
\centerline {within a Two-Band Extended Peierls-Hubbard Model}

\centerline {Y. Anusooya$^{1}$, Swapan K. Pati$^{1}$ and S. Ramasesha$^{1,2}$}

\centerline {$^1$ Solid State and Structural Chemistry Unit} 
\centerline {Indian Institute of Science, Bangalore 560 012, India}
\centerline {$^2$ Jawaharlal Nehru Center for Advanced Scientific Research}
\centerline {Jakkur Campus, Bangalore 560 064, India}
\date{\today}
\maketitle

\begin{abstract}
The phase diagram of halogen-bridged mixed-valent metal
complexes ($MX$) has been studied using a two-band extended Peierls-Hubbard
model employing the recently developed Density Matrix Renormalization Group
method.
We present the energies, charge and spin density  distributions,
bond orders, charge-charge and spin-spin correlations, in the ground
state for different 
parameters of the model. The effect of bond alternation and site-diagonal 
distortion 
on the ground state properties are considered in detail. We observe
that the site-diagonal distortion plays a significant role in deciding the
nature of the ground state of the system. 
We find that while
the $CDW$ and $BOW$ phases can coexist, the $CDW$ and $SDW$ phases are 
exclusive in most of the cases. 
We have also studied the doped $MX$ chains both with and without
bond alternation and site-diagonal distortion in the $CDW$ as well as $SDW$ 
regimes. We find that the additional charge in the  
polarons and bipolarons for hole doping are confined to a few sites,
in the presence of bond alternation and site-diagonal distortion. For
electron
doping, we find that the additional charge(s) is(are) smeared over the entire
chain length and although  energetics imply a disproportionation of the 
negatively charged bipolaron, the charge and spin density distributions
do not reflect it. Positively
charged bipolaron disproportionates into two polarons in the $SDW$ region.
There is also bond order evidence for compression of bond length for the
positively
charged polaronic and bipolaronic systems and an elongation of the bonds
for systems with negatively charged polarons and bipolarons.
\end{abstract}

\vspace{0.8cm}]

\narrowtext
\section{Introduction}
The halogen-bridged mixed-valent metal complexes (HMMC) are 
quasi-one-dimensional chains that exhibit both 
Peierls distortion and mixed valency. This is attributed to the presence 
of strong electron-electron interactions as well as strong electron-lattice 
interactions. Besides, the degeneracy of the ground state 
of the HMMC chains supports solitonic excitations as in polyacetylenes. 
These aspects of HMMCs have resulted in considerable theoretical 
and experimental focus in recent times\cite{tin,nasu89,hua,yuki,kuro,sun1}.

The HMMCs are composed of transition metal ($M$) ions which are bridged by 
halide ($X$) ions. Each metal ion is surrounded by four monodentate ligand 
molecules such as ethylamine ($L$), or two bidentate ligand molecules
such as ethylenediamine, cyclohexanediamine ($L_{2}$), etc. Symbolically,
HMMCs can be represented as $[M^{3-\rho} L_{4}][M^{3+\rho}X_{2}L_{4}]Y_{4}$,
where $M$ can be $Pt$, $Pd$ or $Ni$ and $X$ can be $Cl$, $Br$ or $I$; 
$\rho $ denotes the deviation of the metal valency from the average value 
of $+3$; and $Y$ is a counter ion such as $X^{-}$ or $ClO^{-}_{4}$ for 
charge neutrality. The $d_{z^2}$ orbital on the metal-ion is singly occupied 
when its oxidation state  is $+3$. Along the $M-X$ backbone, 
the electrons are delocalized due to the overlap of the $d_{z^2}(M)$ and
$p_{z}(X)$ orbitals. If the electron-electron interactions are weak compared 
to the electron-phonon interactions, the diagonal electron-lattice interactions
would dominate. This would result in a Jahn-Teller distortion of
opposite phases at successive metal-ion sites. The metal-ion site at which
the $d_{z^2}$ orbital is stabilized would be doubly occupied while
that at the adjacent metal site would be empty leading to a $CDW$ state.
In the opposite limit, the strong electron-electron interactions
force single occupancy of the metal $d_{z^2}$ orbitals and the chain
would be undistorted as there is no electronic stabilization associated
with the distortion due to single occupancy of the metal orbitals. 
In this limit, a $SDW$ state would result. Platinum being a $5d$ system, 
the $d-$orbitals are more diffused resulting in weaker electron-electron 
interactions and indeed, the broken symmetry state observed in the $Pt$ 
complexes is usually a $CDW$ state. At the other end is the nickel system 
with compact $d-$orbitals and one usually observes $SDW$ states in the
$Ni$ complexes. The amplitude of these $CDW$ or $SDW$ distortions 
can be tuned continuously by changing the metal ion, the halide ion,
the ligand, or the counter ions. 

The $CDW$ ground state in these systems has two degenerate 
configurations and hence there is a possibility of soliton-like 
excitations, besides polaronic excitations. The solitonic states in the $MX$
chains are more localized and are believed to be longer lived than 
the solitons in polyacetylene chains. In the halogen bridged mixed-valent
platinum   complexes (HMPC)\cite{kuro}, 
evidence for midgap absorption, associated with solitons, 
comes from the high pressure studies of
the optical spectra, wherein a band at half the
charge-transfer excitation energy is found on application of pressure.
The $IR$ and Raman studies of a series of HMMCs with decreasing
metal-metal distances have also been studied to simulate pressure
and determine several microscopic parameters essential for 
theoretical modelling of the $PtCl$ chains\cite{albe}.
The electrical conductivity and electron spin resonance ($esr$)
studies\cite{haru} of halogen doped
HMPC systems show that for low-doping concentrations, polarons 
are formed which lead to enhanced conductivity and the charge 
carriers are found to have a spin. At increased dopant levels, the
$esr$ intensity reduces although the conductivity increases. Furthermore,
optical absorption studies show the appearance of peaks below the
optical gap. Hence, these studies suggest that two positively charged   
polarons yield either  two positively charged solitons or a bipolaron,
both of which do not have a spin. Photoinduced $IR$ absorption 
studies\cite{okam} on HMPC systems with weak 
interchain interactions have shown evidence for photogenerated solitonic
states, besides polaronic states.
For large interchain coupling, the energy of formation of solitons is 
high and the midgap absorption band in such systems was absent, although
the polaronic absorptions could be observed in these systems.

The $MX$ chains were modelled by using a half-filled 
single-band Hubbard-Peierls model including nearest-neighbour 
electron-electron interactions by many authors
\cite{yuki,nasu83,nasu84,nasu85}.
Nasu, in the mean-field limit, obtained a phase-diagram for the 
nature of the ground state, in the parameter space of $U$, the on-site
electron correlation strength, $V$, the nearest neighbour electron-electron 
interaction parameter and $S$, the strength of site-diagonal electron-phonon 
coupling. The mean-field phase diagram showed regions where, $CDW$ and $SDW$ 
ground states exist as well as regions of coexistence of these two phases. 
It was further shown, within the mean field theory for electrons and
an adiabatic approximation for the phonons that the origin of the 
photoinduced absorption was a distant hole-polaron or an electron-polaron
pair in the excited state of the $MX$ chain.
However, this model apart from the approximation in which it was solved
for, was quite inadequate due to the neglect of the $p_{z}$ 
orbitals on the halogen sites. Ichinose\cite{ichi} mapped the model to an 
anisotropic spin chain, in the limit of small on-site correlations and 
adiabatic electron-phonon coupling to describe the topological excitations of 
the $MX$ chain. Onodera\cite{onod} considered the continuum limit of the 
Ichinose model and showed that it leads to the Takayama-Lin-Liu-Maki model, 
which is also the continuum limit of the discrete Su-Schreiffer-Heeger 
(SSH) model. He showed that the $MX$ chains can support solitons as in 
polyacetylene. A model similar to the SSH model was studied by Baeriswyl 
and Bishop\cite{baer} who showed the existence of a charge-transfer state 
in the limit of strong electron-phonon interactions. The intrinsic defect
 states such as the polarons, bipolarons and solitons, in this limit, were 
observed to be strongly localized. 

Gammel {\it et al} first modelled the $MX$ chains by employing a two-band 
$U-V$ model (consisting of the metal $d_{z^2}$ orbital and the halogen $p_z$
orbital) at ${3 \over 4}^{th}$ filling. This model was studied by
them in different approximations. In the period 4 case, they observed that
the $BOW$ phase exists only in a very small region near the site energies of
 $M$ and $X$ being zero, unlike in the single-band model where the $BOW$
 is found in a wide range of parameter values. Moreover, they characterized
 the lowest state in that region to be of mixed $CDW/BOW$ character.
 They also predicted long period
charge density wave ground states in the system, from an analysis of the
model in the localized limit. They studied the topological excitations
of the model treating the lattice in adiabatic approximation and
the electron-electron interactions in the Hartree-Fock (both restricted and
unrestricted) approximation for various parameter 
values to characterize these excitations in different systems. For small
$MX$ chains, they went beyond the Hartree-Fock approximation and
studied the properties of the chain by employing exact diagonalization 
methods. They also studied the model including the phonon dynamics 
but treating the electron-electron interactions in the mean field limit.
For small Hubbard interaction strengths, perturbation theory was employed
to study the model.  Huang and Bishop\cite{hua} studied the two-band 
model both in the mean-field and random phase approximations to 
study the lattice- and spin-polaronic defects in $Ni$ complexes. 
They found  relative lattice-distortion around the defect center 
besides the charge or spin disproportionation. The effect of 
interchain interactions on the nature of the ground state and also on
the energy gaps in the system were studied by including them
self-consistently
in finite $MX$ chain calculations, within a two-band model\cite{yam}. The 
effect of interchain interactions on the stability of nonlinear lattice 
relaxation was considered by Mishima\cite{mish} in the mean-field 
approximation within the one-band extended Hubbard-Peierls model.
Sun {\it et al} \cite{sun2} employed a one-band model and in the mean-field
approximation showed  that the electron-electron interaction
reduces the $CDW$ gap in $MX$ complexes. There is also an all-electron
local density approximation calculation for $MX$ chains which
focuses on the band-gap, dimerization and $SDW$ instabilities in these
compounds\cite{anis}.

All the studies so far carried out on the $MX$ chain systems 
suffer from the disadvantage that they treat
electron-electron interactions in the mean-field approximation,
except in the case of small chains where model exact solutions
are obtained. The exact studies on small chains are  often inconclusive
due to finite size effects. However, the recently developed Density
Matrix Renormalization Group (DMRG) method has proved to be very accurate
for quasi-one-dimensional systems\cite{srwh}.  In this paper,
we report results of our extensive investigations of the $MX$ chain
systems employing the DMRG method. We have
studied the $MX$ chains with upto 70 sites (35 $MX$ units), employing
the two-band extended Peierls-Hubbard model. We have studied the neutral
as well as
charged $MX$ chains  to understand the properties of ground 
state as well as the photogenerated gap states for many values of the model
parameters. Besides energies, we have studied the charge and spin 
correlations in the system, charge and spin densities as well as
bond orders to properly characterize the  ground states in different
regions of the parameter space.
The paper is organized as follows. In the next section we introduce the
model Hamiltonian and the DMRG method as applied to the $MX$ chains. In the
third section, we discuss results for the ground state of the neutral
and doped systems. 

\section{Model Hamiltonian and the DMRG Method} 
We have studied the HMMC systems employing the $U-V-\delta$ model. 
The  Hamiltonian of this model, ${\hat H}$, for the 
metal-halogen chain can be 
written as a sum of the noninteracting term, ${\hat H_0}$, which includes
the renormalized static electron-lattice interactions and an 
electron-electron interaction term, ${\hat H}_1$, given by
\begin{eqnarray}
{\hat H} & = & {\hat H}_{0}+{\hat H}_{1}  \\
{\hat H}_{0} & = & \sum_{i=1}^{N} \sum_{\sigma} t_{i}
[a^{\dagger}_{X,i\sigma }a_{M,i\sigma }+ a^{\dagger }_{M,i\sigma } 
a_{X,i+1\sigma }+H.C.]  \nonumber \\ 
&  & +\sum_{i=1}^{N} \sum_{\sigma } [\epsilon_{M,i}a^{\dagger}_{M,i\sigma }
a_{M,i\sigma} + \epsilon_{X,i}a^{\dagger}_{X,i\sigma }a_{X,i\sigma} ] 
\end{eqnarray}

\begin{eqnarray}
{\hat H}_{1} & = & \sum_{i=1}^{N} U_{M} \frac {{\hat n}_{M,i}
({\hat n}_{M,i}-1)}{2}+
\sum_{i=1}^{N}U_{X} \frac {{\hat n}_{X,i}({\hat n}_{X,i}-1)}{2} \nonumber \\
&  & +\sum_{i=1}^{N}V_{i}
[{\hat n}_{M,i}{\hat n}_{X,i+1}+{\hat n}_{X,i}{\hat n}_{M,i}]
\end{eqnarray}

\noindent
where $t_{i}$ is $t (1- (-1)^{(i)} \delta)$.
The summations run over all the $N$  $MX$ pairs and the upper limit
of the summation is chosen to reflect  open boundary condition corresponding 
to a chain.  $a^{\dagger}_{X,i\sigma}$  ($a^{\dagger}_{M,i\sigma}$) 
 creates an electron with spin $\sigma $ in the halogen (metal)
orbital in the $i^{th}$ unit cell and  $a_{X, i\sigma}$ ($a_{M, i\sigma}$) 
is the adjoint of the corresponding creation operator.  The operators 
$\hat n_{X,i}$ (${\hat n}_{M,i}$)  are the number operators for the halogen 
(metal) orbital in the $i^{th}$ unit cell. 
$\epsilon_{M,i}$ ($\epsilon_{X,i}$)  is the site energy of the metal (halogen)
orbital in the $i^{th}$ unit cell.  $U_{M}$ ($U_{X}$) is
the on-site electron-electron repulsion parameter for the metal (halogen) 
orbital. The nearest neighbour electron-electron interaction terms $V_{i}$s
are calculated using Ohno\cite{ohno} interpolation scheme,
\begin{equation}
V_{i}=14.397[28.794/(U_{M}+U_{X})^{2}+ r^{2}]^{-1/2}
\end{equation}
where $r$ is the distance between the nearest neighbours of the $MX$ chain.
The distance $r$ between the pairs depends on the alternation parameter
$\delta$.
All the parameters are defined in units of the uniform transfer integral
$t$.

We have employed the Density Matrix Renormalization Group (DMRG) method
to obtain the ground state properties of the above Hamiltonian for large
$N(\geq 35)$ where $N$ is the number of $MX$ pairs. In the DMRG method 
for the $MX$ chains, we start with two $MX$ units(4 sites) and obtain 
the ground state of this cluster with six electrons corresponding to 
${3 \over 4}^{th}$ filling by an exact diagonalization procedure. We now 
imagine the chain to be built up of two halves, namely the left-half and
the right-half. We construct the reduced many-body density matrix of the 
left-half, $\rho_{0,L}^{(2)}$ , in the basis of the Fock space states of 
the left half of the chain from the ground state eigenfunction by
 integrating over the Fock space states on the right-half as, 

\begin{equation}     
({\hat \rho}^{(2)}_{0,L})_{\mu \nu} = \sum_{\mu^{\prime}}
 C_{\mu \mu^{\prime}} C_{\nu \mu^{\prime }} 
\end{equation}
\noindent
where $|\mu>$ and $|\nu>$ are the Fock space states of the left-half
chain and $|\mu^{\prime}>$ the Fock space states of the  right-half chain.
$C_{\mu \mu^{\prime }}$  is the coefficient
 associated with  direct product functions $|\mu>$ and $|\mu^{\prime}>$ 
 in the ground state eigenfunction. The dimensionality of the Fock-space 
$|\mu>$ for a  system consisting of $n$ units is $l=4^{2n}$. The
density matrix is simultaneously block diagonal in both the particle-number 
sector and in the $M^{L}_{s}$ sector where $M^{L}_{s}$ is the z-component
of the total spin of the left-half block.
We take advantage of this while diagonalizing the density matrix by
diagonalizing each of the   blocks independently. This also allows us to 
label each density matrix eigenvectors by the particle-number, $p_{L}$
besides $M^{L}_{s}$. After diagonalization, the Fock space on the left
is truncated by retaining only  $m$ of the density matrix
eigenstates corresponding to the $m$ highest density matrix eigenvalues.
If we had retained all the $l$ density matrix eigenvectors
 to serve as basis
functions of the Fock space of the left-half, we would have merely effected 
an unitary transformation of the basis functions. The  $l \times l$ 
Hamiltonian matrix ${\hat H}_{L}(n)$   for the left 
 part of the chain are obtained in the basis of the 
Fock space states. This matrix ${\hat H}_{L}(n)$   is 
renormalized using the matrix $\hat O_{L}(n)$  whose columns
are the $m$ eigenvectors of the corresponding $l\times l$ density matrix.
Thus the transformation matrix ${\hat O}_{L}(n)$ is a ($m \times l$) matrix. 
The renormalized  Hamiltonian 
matrix  $\tilde H_{L}(n)$ is  given by
\begin{equation}
{\tilde H}_{L}(n)= {\hat O}^{\dagger}_{L}(n) {\hat H}_{L}(n) {\hat O}_{L}(n)
\end{equation}
\noindent
The renormalized left Hamiltonian matrix is  now an $m \times m $ matrix
representation of the left half-Hamiltonain  in the basis of the density 
matrix eigenvectors. The operators $a^{\dagger}_{M,i}$  and 
$a^{\dagger}_{X,i}$ and ${\hat n}_{M,i}$ and ${\hat n}_{X,i}$ 
corresponding to each site  in the left part of the chain are also
obtained as matrices in the basis of
the Fock space $|\mu>$  and are later renormalized to obtain
renormalized
matrices in the basis of the eigenvectors of the density
matrix of the corresponding half-chains, in a manner similar to 
the construction of ${\tilde H_{L}}(n)$. The density matrix,  the transformation
matrix ${\hat O}_{R}(n)$, the renormalized Hamiltonian matrix
${\tilde H_{R}}(n)$,  for the right part as well as the renormalized 
second-quantized site operators for the right part are all obtained
analogously. Unlike in the calculations 
involving spin chains and Hubbard chains\cite{skp}, the $MX$ chains
 do not have
the reflection symmetry and all the quantities should be calculated 
separately for the right and the left halves of the chain. For this 
reason, we also cannot iterate the DMRG procedure to obtain self-consistent
density matrices for fragments of different sizes of the targetted
$MX$ chain, a method that is usually employed to obtain more accurate 
properties for a chain of given length, in systems with reflection 
symmetry\cite{srwh}.

To get the Hamiltonian for the system with ($n+1$) unit cells, a $MX$ unit
is added in the middle of the chain.  The Hilbert space of the new
Hamiltonian matrix corresponding to  ($n+1$) unit cells is the direct
product of $m$ states, $|\mu>$ from left block and $|\mu^{\prime}>$
from right block  and 4 states, ${|c>}$ or ${|c^{\prime}>}$
(corresponding to $|0>$, $|\downarrow>$, $|\uparrow>$, and
$|\uparrow \downarrow>$ configuration at
the new site) from each of the newly added unit cell, with the
restriction that the total $M_{s}$ value for the full chain is equal to 
the desired value  and that the total system is ${3 \over 4}^{th}$ filled.

The Hamiltonian for ($n+1$) unit cell system can be written as
\begin{eqnarray}
{\lefteqn{{\hat H}(n+1)  = {\tilde H_{L}}(n)+ {\tilde  H}_{R}(n)  
+ {\hat n}_{c} \epsilon_{c} + {\hat n}_{c^{\prime}} \epsilon_{c^{\prime}} 
\nonumber }} \\
& & + \frac{U_{c}}{2} {\hat n}_{c} ({\hat n}_{c}-1) 
+ \frac {U_{c^{\prime}}} {2} {\hat n}_{c^{\prime}} ({\hat n}_{c^{\prime}}-1)
  \nonumber \\
& & +t_{n}[{\tilde a}^{\dagger}_{L}(n)a_{c}+h.c.]+ 
t_{n+1}[a^{\dagger}_{c}a_{c^{\prime}}+h.c.] 
\nonumber \\
& & + t_{n+1}[a^{\dagger}_{c^{\prime}}{\tilde a}_{R}(n)+h.c.]
+ V_{n}{\tilde n}_{L}(n) {\hat n}_{c} +
 V_{n+1}{\hat n}_{c} {\hat n}_{c^{\prime}} \nonumber \\
& & + V_{n+1}{\hat n}_{c^{\prime}} {\tilde n}_{R}(n)
\end{eqnarray}
\noindent
where the operators ${\tilde a}^{\dagger}_{L}(n)$, 
${\tilde a}^{\dagger}_{R}(n)$
and their adjoints as well as ${\tilde n}_{L}(n)$ ,
 ${\tilde  n}_{R}(n)$ 
are the renormalized operators expressed in the truncated density matrix 
eigenvector basis. The operators, ${\hat a}^{\dagger}_{c}$ 
( ${\hat a}^{\dagger}_{c^{\prime}}$) and their adjoints as well as
${\hat n}_{c}$ (${\hat n}_{c^{\prime}}$) are expressed as matrices in the 
Fock space basis. The matrix
 representation of the Hamiltonian ${\hat H} (n+1)$ in the direct products 
 basis is obtained as appropriate
direct product of the operators occuring in the Hamiltonian.

The eigenvalues and eigenvectors for this ($n+1$) unit cell
are obtained and  the reduced density matrices for the left and right
half of the chain, each with ($n+1$) sites are constructed from the ground 
state eigenfunction. In the next iteration,
the procedure is repeated by adding a $XM$ unit in the middle of the chain.
Alternately we have to add $MX$ and $XM$ units in the middle of the chain so
that the successive sites of the full chain at any iteration is not 
occupied by like ions.

  We have optimized the DMRG cut-off, $m$, by comparing the ground state
energy per $MX$ unit for different cut-offs for chain lengths ranging 
between 25 to 35 $MX$ units. We have presented these energies in table 1.
We find that a value of $m=80$ is quite satisfactory. We have used the
DMRG cut-off, $m=80$, in all our calculations. The dimensionality
of the Hilbert space corresponding to $M_s = 0$ (4n sites system)
or $0.5$ (4n+2 sites system) and $N_{e} = {3 \over 2} N$
varies in the range $6400$ to $7000$, depending upon the model  parameters, 
for this value of $m$.  The resulting Hamiltonian matrix is very sparse. 
The total number of nonzero matrix elements are $\approx$ 350000. 
We exploit the sparseness of the Hamiltonian matrix to reduce the
storage requirement as well as $CPU$ requirement by avoiding
doing arithmetic with zeroes.  We have used the Davidson algorithm
for symmetric Hamiltonian matrix to get the lowest few eigenvalues.
Davidson algorithm,
which is a hybrid  of coordinate relaxation method and Lanczos method
has been widely used in  quantum chemical computations and is known to 
be both robust and rapidly convergent. The properties of the chain are
computed by using the renormalized matrices for the site operators 
and product operators (for bond order calculations) after reaching the
desired length of the chain.

We have compared the DMRG results with results from exact diaganolization
for small systems. The exact diagonalization of the model has been carried 
out using the Diagrammatic Valence Bond (DVB) 
method. The DVB method exploits the total spin conservation property
of the model Hamiltonians and employs a valence bond (VB) basis in which
the Hamiltonian is represented as a matrix\cite{srso}. As the VB basis
 is nonorthogonal, the resulting Hamiltonian matrix is nonsymmetric.
We have used Rettrup's algorithm  which is similar to Davidson's algorithm
for symmetric matrices, to obtain the lowest few eigenvalues 
and eigenvectors of the nonsymmetric matrix\cite{ret}. 
This exact diagonalization calculation is quite straightforward
for chains of upto  seven $MX$ units.

Table 2 compares the exact ground state energies with the DMRG ground
state energies for two different parameter sets of the Hamiltonian.
 Also given for comparison
are the dimensionalities of the Hilbert space for $M_s=0$ in the VB scheme
(exact) with the $M_{s}=0$ in the DMRG scheme (with cut-off). The DMRG scheme is found to 
reproduce accurately the  ground state energies where such comparison
is possible. We have also compared the DMRG properties such as charge 
density, $<{\hat n}_{i}>$, spin density, $<S^{z}_{i}>$ and bond order,
 $-{1 \over 2} <a^{\dagger}_{i \alpha}a_{j \alpha}+
 a^{\dagger}_{i \beta}a_{j\beta}+h.c.>$  
with exact properties (table 3 and 4). We find that the DMRG ground state
properties are in very good agreement with the exact properties.

\section{Results and Discussion}
We have studied the ground state of the $MX$ chains 
in different regions of parameter space to study the phase transformation
from
$CDW$ phase to $SDW$ phase.  The parameters $U_{M}$, $U_{X}$, and
$\epsilon_M$, $\epsilon_{X}$ characterize the metal ion and the halide ion.
The orbital
energy of the
halide ion, $\epsilon_{X}$, is specified relative to the orbital energy
of the
corresponding metal ion of the uniform $MX$ chain. $\epsilon_{X}$ is always
negative reflecting the larger electronegativity of the halides compared to
the metal ions. In the halide series, larger negative $\epsilon_{X}$ 
represents chloride while the least negative $\epsilon_{X}$ represents 
iodide reflecting the electronegativity variations in the halogen group.
The  on-site repulsion parameter $U_{X}$ is positive and decreases as we go 
down the series from 
$Cl^{-}$ to $I^{-}$. The Hubbard parameter for the metal ion, $U_M$, decreases as we go
from the $I$ row  transition elements to the $III$ row transition elements. 
The parameters $U_{M}$, $U_{X}$ and $\epsilon_{X}$ are varied from
$U_{M}=2.5t$,
$U_{X}=t$, $\epsilon_{X}=-2t$ to   $U_{M}=1.5t$,
$U_{X}=0.5t$, $\epsilon_{X}=-t$ corresponding to the $MX$ pairs $NiCl$ to $PtI$
respectively. The $\epsilon_{M}$ values depend upon the strength of the 
diagonal electron-lattice coupling and so does the alternation $\delta$ in
the transfer integrals. The  coupling constants 
for the diagonal and off-diagonal couplings are assumed to be independent.
Accordingly, we independently vary the
transfer integrals as well as the site energy at the metal site,
$\epsilon_M$. 
This is one of the
crucial differences between a polyene chain and the $MX$ chain.
In the former,
the site-diagonal electron-phonon coupling is taken to be zero while
in the $MX$ chains it is nonzero by virtue of the crystal-field environment
provided by the halide ions surrounding the metal ions. The dimerization 
parameter, $\delta$, has been varied between 0.0 and 0.2. 
 In what follows, we first 
discuss the results of our study of $MX$ chains at ${3 \over 4}^{th}$ 
filling and then discuss our results for these chains with one and two 
excess (fewer) electrons. 
\subsection{\bf $MX$ chains at ${3 \over 4}^{th}$ filling}

In fig.1 we present the dependence of the ground state energy per
$MX$ unit (${\epsilon}_{MX}$)  of the $MX$ chains for different values
of  $\delta$ for one set of parameters. The convergence to the
infinite chain value is monotonic and from below. We have defined
${\epsilon}_{MX}$ as half the total energy difference between successive
iterations which differ by two $MX$ units. This definition corresponds
to the energy of an embedded $MX$ unit and is akin to the rings. It
is well known that the energy per site of Hubbard, extended Hubbard
as well as spin rings converges to the limiting value from below \cite{soos}. 
In table 5, we have shown the dependence of
fractional stabilization of the $MX$ chain on introducing alternation
for several sets of parameters. We find that
the alternation lowers ${\epsilon}_{MX}$ in all the cases we have
studied but the extent of stabilization is rather small and insensitive
to variations in $U_M$ and $\epsilon_X$ when the diagonal electron-phonon 
coupling is neglected. While from finite chain studies it is not 
possible to reliably deduce
whether for a given set of parameters, the distortion of the chain is
unconditional (independent of the lattice stiffness), our results indicate 
the dominant role of diagonal electron-phonon interactions in 
determining the extent of bond alternation. In fact, in systems where
bond alternation is indeed found, the magnitude of the alternation
is very large.
                                                      
In fig.2 we present the charge densities at the metal and halide sites in
the alternating ($\delta = 0.1$) $MX$ chains without diagonal distortions
($\epsilon_M = 0$)
for two extremal values of $U_M$ and $\epsilon_X$. We find that the charge 
densities at the metal sites are very nearly uniform in both cases. For small
values of $\epsilon_X$ and $U_M$, the charge density in the metal orbital is 
larger at $\approx 1.2 \pm 0.08$ while it is more uniform with values 
in the range $\approx 1.06 \pm 0.04$ for large $U_M$ and large $\epsilon_X$.
The charge densities at the halide sites are uniform and closer to 2 electrons
when $\epsilon_X$ and $U_M$ are large. While the alternation in the 
transfer integral along the chain seems to promote mixed valency, large
electron-electron repulsion at the metal site and large site energy
of the halogen orbital has the effect of suppressing mixed valency. This
is in conformity with experiments wherein mixed valency is found in
$MX$ chains with heavier transition elements as well as heavier halogen
atoms. Increasing the alternation in transfer integral does not change the
picture significantly. There is a slight increase in the amplitude of
the charge density wave in the most favourable case we have studied,
corresponding to $U_M = 1.5t$ and $\epsilon_X = -t$ (fig.3).

In fig.4, we show the plot of charge density at metal sites in the presence
of diagonal distortion ($\epsilon_M \ne 0.0$) for alternation 
$\delta = 0.1$
for the two extremal cases we have studied, namely, large $U_M$, large 
$\epsilon_X$ and small $U_M$ and small $\epsilon_X$ for one particular 
value of $\epsilon_M$. We see a dramatic change in the charge density
distribution in both cases. In the favourable case, the 
disproportionation of the metal ion in 3+ oxidation state into 2+
and 4+ oxidation states is almost complete. While even in our least
favourable case, the amplitude of the $CDW$ is quite significant. In the
latter case increasing $\epsilon_M$ increases the amplitude rapidly.
This result underlines the importance of diagonal distortion in
producing a $CDW$ ground state.

The variation in bond order along the $MX$ chain is plotted for
several values of the parameters in fig.5 only for the left half of the 
chain. The amplitude of the
bond order wave ($BOW$) behaves similar to the amplitude of the $CDW$
.
The diagonal distortion has a strong effect on the $BOW$ amplitude. 
Even in the most unfavourable case of $U_M = 2.5t$ and $\epsilon_X = -2t$,
the $BOW$ picks up sufficient amplitude for the site diagonal distortion
we have considered. The earlier prediction of Gammel {\it et al}\cite{tin}
that a BOW cannot exist for negative halide site energy with site 
diagonal distortion lowering the metal-ion site energy is not borne 
out by our calculations.

The spin density distribution (fig.6) shows a trend opposite to what is
observed with $CDW$ and $BOW$ instabilities. For large $U_M$ and
large $\epsilon_X$,
in the uniform $MX$ chain, the $SDW$ amplitude is fairly large.
Introducing off-diagonal alternation reduces the amplitude, although
the alternation in the spin density still exists. However, on introducing
diagonal distortion, all the metal sites become completely nonmagnetic.
This behaviour shows that when the $CDW/BOW$ amplitude is large the
amplitude of the $SDW$ is small. Fig.7 brings out this trend by comparing
the charge and spin densities at metal sites for different values of the
site-diagonal distortion parameter $\epsilon_M$. It is also interesting
to note that for one set of parameter values, the $CDW$ and the $SDW$
phases coexist (fig.7 ii).

We have also characterized the ground state in various regimes of the
parameters by
studying the spin-spin and charge-charge correlation functions.
Although these correlation functions have been computed for open chains,
they can be Fourier transformed, if one assumes that the  correlations
in the interior of the open chain are close to what would be seen
in a ring. This assumption was first made by Affleck {\it et al}\cite{ian} 
to obtain the structure factors from open chain DMRG calculations of
spin systems. In our calculations, we have discarded the last three
unit cells on either ends of the chain and have assumed the correlations
to have a reflection symmetry about the middle bond. This would 
enable us to Fourier transform the correlation functions. 

In fig.8, we show the spin-structure factor, $S(q)$, for various values of 
the model parameters. In the $CDW$ phase which corresponds to small $U_M$,
small $\epsilon_X$ and nonzero site-diagonal distortion and alternation,
we find that $S(q)$ is very small. However, for large $U_M$, large
$\epsilon_X$, zero site-diagonal distortion but with nonzero $\delta$,
the structure-factor is large and peaks at $q = {\pi}$. This result
reflects the spin ordering of the ground state. The uniform structure
factor in fig.8a and fig.8d confirms the ground state to be in $CDW$ phase. This
result is also consistent with the charge and spin density and the bond
 order
data discussed above. 
In fig.9 we show the structure factor corresponding
to the charge-charge correlation function. Here again, 
for large $U_M$ and
large negative $\epsilon_X$, the structure factor is almost uniform
and does not show any pronounced peaks. However, for small $U_M$, small
negative $\epsilon_X$ and nonzero site-diagonal distortion, the structure
factor peaks at $\pi$ corresponding to the existence of a $CDW$ phase.
The importance of the diagonal distortion is underscored by the fact
that even for $\delta$, small $U_M$ and small negative $\epsilon_X$, 
the peak at $\pi$ in the strcuture factor though discernible, is not
pronounced.
It is also interesting to note that $\rho(q)$ shows
small oscillations away from the peak at $\pi$ which could be due
to incipient long-wave length $CDW$ distortions which could have
nonzero amplitude in the thermodynamic limit as suggested by Gammel 
{\it et al}\cite{tin}.

\subsection{\bf $MX$ chains marginally away from ${3 \over 4}^{th}$ filling}
The DMRG method for $MX$ chains cannot access the energy levels
that have been studied by optical spectroscopies. The reason being,
there are a large number of low-lying excitations in long $MX$ chains
which intrude while targetting excited states and absence of the
symmetries in open chains rules out the possibility of avoiding
the intruder states. Hence, we have been unable to study the optical
properties of long $MX$ chains by this technique. However, there is
considerable interest in the photogenerated gap states which arise
from the dissociation of the excitons produced in an optical experiment.
These states are typically, the positive and negative polarons and
bipolarons and the charged and neutral solitons of the system. The
DMRG method can easily access the polaronic and bipolaronic states.
In what follows, we present results of the DMRG study of these
species at representative points in the parameter space, namely 
$U_M = 1.5t, \epsilon_X =-t$
and $U_M = 2.5t, \epsilon_X =-2t$ corresponding to the $CDW$ and $SDW$
regimes. These parameters are taken together with $\epsilon_M=0.0$ 
or $t$ and $\delta=0.1, U_X=0.5t$ to explore the importance of 
site-diagonal distortion in the two regimes.

We first discuss the energetics of the  polarons and bipolarons, for
the chosen parameter set. 
In table 6 is given the energy for doping of $MX$ chains
of 35 units with one or two holes and one or two electrons.
 The magnitude of doping energy
increases with increase in strength of electron correlations.
The stabilization energy on doping with one(two) hole(s) is(are)
almost equal in magnitude to the energy required for creating
one(two) electron(s) doped chains respectively. Neither the bond 
alternation
nor the site-diagonal distortion energy at the metal site
have any noticeable influence on the doping energetics. However,
the positively charged bipolaron and the negatively charged bipolaron
 are not placed symmetrically around the ground state in the
energy scale. From the energetics, one can see that at larger
correlation strengths, the positively charged bipolaron is less stable
than the two positively charged polarons. This is also true
for negatively charged bipolaron irrespective of Hubbard $U$. Thus, it
appears from the energetics that both the positively and the
negatively charged bipolarons should dissociate into two polarons.

The definitive proof for the disproportionation of the bipolarons
comes from comparing the charge and spin density distributions of
the bipolarons with those  of the polarons bearing charges of the same sign.
The charge densities 
at the metal site for the polarons and bipolarons are shown in fig.10.
The polaron charge densities for (I) $U_M=1.5t$,
$\epsilon_M=0.0$,  $U_X=0.5t$ and $\epsilon_X=-t$ are shown in fig.10a, 
and for (II) $U_M=2.5t$,
$\epsilon_M=0.0$, $U_X=0.5t$ and $\epsilon_X=-2t$ are shown in fig.10c. 
The data for
bipolarons for the parameter set (I) are shown in fig.10b and for the set
(II) are shown in fig.10d. For the parameter set I (fig.10a and b), 
the additional charge
is uniformly distributed over the entire chain for the (i) positively charged
polaron/bipolaron, (ii) the neutral chain and (iii) for the negatively
charged polaron/bipolaron. For the second set of parameters,
{\it i.e.} at large $U_M$, we observe more localized
charge distribution for both the positively charged polarons and positively
charged bipolarons. We also observe two broad peaks (fig.10d) in the charge
distribution of the positively charged bipolaron which is indicative of
disproportionation of the positively charged bipolaron into two positively 
charged
polarons. However, in the case of the negatively charged polarons and
bipolarons, the Hubbard $U$ prevents the localization of charge. 
An earlier mean-field study \cite{tin} found the negatively charged polaron
and bipolaron to be more localized than the positively charged polaron
and bipolaron. Our DMRG results correspond to an on-site halide 
repulsion parameter $U_X$, which is smaller than the metal on-site
repulsion parameter, $U_M$ and our study should have enhanced this
difference between the hole-defects and the electron-defects predicted by
the mean-field analysis. It appears, therefore,
that the mean-field approximation gives wrong trends for charge distributions
of the defects. On physical grounds, one should expect that the on-site
repulsions spread out the excess negative charge more than excess positive
charge.

The evidence for the disproportionation of the positively charged
bipolaron into two polarons is more pronounced when alternation in the
chain is introduced (fig.11). The charge density distribution
of the polaron (fig.11a) shows a single hump while that for the bipolaron
(fig.11b) shows two humps. The hole charge density is mostly confined
to one sublattice of the metal ion, the one in which the metal-halogen
bond is shorter accommodates the excess charge. The halogen charge
density distribution is not affected significantly by doping. 
The spin density distribution shows
the disproportionation more clearly  as seen from the two separate 
envelops for the spin density in the bipolaron (fig.11d) compared 
to a single envelop in the polaron spin density distribution 
(fig.11c). This break-up of
the bipolaron is observed only for hole doping. 

The effect of site-diagonal distortion on the disproportionation is
very dramatic. We compare the charge density distribution for
the positively charged  polarons and bipolarons with (fig.12a) and without
(fig.12b) site diagonal distortion. In both the cases, the localization 
of excess charge is confined to only one sublattice. In the system with 
site-diagonal distortion, 
the sublattice with nonuniform charge density is on the metal site for
which the metal-halogen bond is long, corresponding to a negative
$\epsilon_M$ while in the absence of site-diagonal distortion,
these metal-sites have uniform charge distribution. 
This is seen
as a change over in the charge density humps from the upper
envelop in fig.12a to the lower envelop in fig.12b. 
The disproportionation of the positively charged bipolaron is again found only
for systems with large on-site repulsions, $U_M$.

The difference between $MX$ chains at ${3 \over 4}^{th}$ filling
and $MX$ chains with one- and two-hole dopings, can be seen clearly, 
if the difference in charge density between corresponding metal sites
of the neutral and doped chains is plotted as a function of site
number. We show in fig.13 these difference plots for systems with
site-diagonal distortion for weak and strong correlation cases.
The envelop of the charge density distribution of the 
polaron and the bipolaron show a single peak
at small correlation strengths, while the same for strong correlations
exhibits two distinct peaks for the bipolaron (fig.13a, iv). A similar
behaviour is also found in the spin density distribution (fig.13b, iv).

The negatively charged bipolaron does not disproportionate even
upon introducing the site-diagonal distortions. In fig.14, we show
for small and large $U_M$ values, the charge density (fig.14a) and spin
density (fig.14b) distributions at the metal sites. Apart from 
exhibiting mixed valency, the charge and spin density distributions 
are uniform on each sublattice for the negatively charged bipolarons.

The bond order distributions in the negatively and positively charged 
bipolarons are almost similar to what is found in neutral chains. 
These are shown in fig.15 only for the left half of the chain.
There is a tendency for the negatively charged bipolarons towards  the
elongation of the bonds (as seen from smaller bond orders in the middle of the
chain), while the positively charged bipolarons have an opposite tendency
 {\it i.e.}, towards
bond length contraction. This agrees with an earlier study of the 
lattice distortions of doped $MX$ chains \cite{hua}. However, these marginal
differences in the bond order variations are reduced on introducing
site-diagonal distortions. The essential difference between the positively
and negatively charged bipolaron lies in the disproportionation of 
the former into positively charged polarons in the strong 
correlation limit, in the
presence of alternation and site diagonal distortion.

  In summary, we have studied the phase diagram of $MX$ chains within a two-band
extended Peierls-Hubbard model employing the density matrix renormalization 
group method. We find that  the site energy associated with site-diagonal 
distortion  is 
the single most important parameter for the transition from a $SDW$ phase to 
a $CDW$ phase in the 
ground state of the system. The variation of other parameters, such as, site energy of the
halide site, on-site Hubbard $U$ of the halide ion and 
bond alternations do not change the
nature of the ground state significantly. On the otherhand, for the doped 
$MX$ chains,
both the bond alternation and site-diagonal distortion play a major role.
For positively doped systems, introduction of bond alternation leads to the
localization of charge and spin densities. In the presence of site-diagonal
distortion, we observe that the positively charged bipolaron 
disproportionates into two positively charged polarons in strong correlation
limit. 
The negatively charged bipolarons
do not show evidence for disproportionation even for the longest chain 
length and for the parameters we have
studied. We also find that there is a contraction
of bond length in the case of positively charged polarons and bipolarons
and elongation for the corresponding negatively charged species.

Acknowledgements:  It's a pleasure to thank Prof. H. R. Krishnamurthy
for many fruitful discussions and Dr. Biswadeb Dutta for system help.

\onecolumn
{\bf Tables} \\

Table 1. Energy per $MX$ unit  for  systems with different DMRG cut-offs, $m$,
for the parameters set $U_{M}=2.5t$, $U_{X}=0.5t$, $\epsilon_{M}=1.5t$, 
$\epsilon_{X}=-2t$ and $\delta=0.1$.
\vspace{0.3cm}
\begin{center}
\begin{tabular}{cccc} \hline 
N of \hspace{0.5cm} & $m = 70 $ \hspace{0.8cm}&  $m = 80 $ \hspace{0.8cm}& $m = 90 $  \\
$MX$ units \hspace{0.5cm} & \hspace{0.8cm}& \hspace{0.8cm} &  \\ \hline
 25  \hspace{0.5cm}&  1.413177 \hspace{0.8cm}&  1.413175 \hspace{0.8cm}&  1.413168 \\
 27 \hspace{0.5cm}&   1.413158 \hspace{0.8cm}&   1.413157 \hspace{0.8cm}&  1.413156 \\
 29 \hspace{0.5cm}&  1.413202 \hspace{0.8cm}&  1.413177 \hspace{0.8cm}& 1.413170 \\
 31 \hspace{0.5cm}&   1.413159 \hspace{0.8cm}&   1.413157 \hspace{0.8cm}&  1.413156 \\
 33 \hspace{0.5cm}&  1.413181 \hspace{0.8cm}&  1.413173  \hspace{0.8cm}& 1.413169 \\
 35  \hspace{0.5cm}& 1.413159 \hspace{0.8cm}&  1.413157 \hspace{0.8cm}& 1.413156 \\
\hline
\end{tabular}
\end{center}
\vspace{0.6cm}

Table 2. Comparison of dimensionality and energy per $MX$ unit  with
 exact calculation. $p$ is the site number and $P_{exact}$ and 
$P_{DMRG}$  are the dimensionality
in the exact and DMRG calculation (with DMRG cut-off $m=80$). $U_{X}=0.5t$, 
$\epsilon_{M}=0$, $\epsilon_{X}=-t$ and $\delta =0.0$. A in the table 
corresponds to $U_{M}=1.5t$ and B to $U_{M}=2.5t$. 
\vspace{0.3cm}
\begin{center}
\begin{tabular}{ccccccccc} \hline
$p$ \hspace{0.5cm}& \multicolumn{4}{c} {A}\hspace{0.7cm} &  
\multicolumn{4} {c} {B} \\ 
 \hline
 \hspace{0.5cm}& $P_{exact}$ \hspace{0.7cm}&  $P_{DMRG}$\hspace{0.7cm} & $E_{exact}$\hspace{0.7cm}& $E_{DMRG}$\hspace{0.7cm}& $P_{exact}$\hspace{0.7cm}& 
$P_{DMRG}$\hspace{0.7cm}&  $E_{exact}$\hspace{0.7cm}& $E_{DMRG}$  \\  \hline 
6 \hspace{0.5cm} & 90\hspace{0.7cm} & 90\hspace{0.7cm} & 4.3834\hspace{0.7cm}    &   4.3834\hspace{0.7cm} &  90\hspace{0.7cm}& 90\hspace{0.7cm} & 13.4129\hspace{0.7cm} &  13.4129 \\
8 \hspace{0.5cm} & 784\hspace{0.7cm} & 784\hspace{0.7cm} &  7.2669\hspace{0.7cm} &    7.2669\hspace{0.7cm}     &  784\hspace{0.7cm} & 784\hspace{0.7cm} &  20.0065\hspace{0.7cm} &  20.0065 \\
10 \hspace{0.5cm} & 5400\hspace{0.7cm} & 3873\hspace{0.7cm} & 10.4504\hspace{0.7cm} &   10.4505\hspace{0.7cm}  &   5400\hspace{0.7cm} & 3930\hspace{0.7cm} & 26.7485\hspace{0.7cm}  & 26.7485 \\
12 \hspace{0.5cm} & 48400\hspace{0.7cm} & 7482\hspace{0.7cm} & 13.4510\hspace{0.7cm} &    13.4527\hspace{0.7cm} &  48400\hspace{0.7cm} & 7687\hspace{0.7cm} & 33.4120\hspace{0.7cm} & 33.4130 \\
14 \hspace{0.5cm} & 364364\hspace{0.7cm} & 6277\hspace{0.7cm} & 16.6227\hspace{0.7cm}  & 16.6229\hspace{0.7cm}  &  364364\hspace{0.7cm} & 6356\hspace{0.7cm} & 40.1543\hspace{0.7cm} & 40.1544\\
\hline 
\end{tabular}
\end{center}
\vspace{0.6cm}

Table 3. Comparison of charge densities  and spin densities for 14 sites chain 
with   exact calculation. $p$ is the site number. A and B correspond to 
the parameter values quoted in table 2.
\vspace{0.3cm}
\begin{center}
\begin{tabular}{ccccccccc} \hline
$p$ \hspace{0.5cm}& \multicolumn{4}{c}{charge density}\hspace{0.7cm}&  \multicolumn{4}{c} 
{spin density}  \\  \hline
\hspace{0.5cm}& \multicolumn{2}{c}{A}\hspace{0.7cm}& \multicolumn{2}{c}{B }\hspace{0.7cm}& 
 \multicolumn{2}{c}{A}\hspace{0.7cm}& \multicolumn{2}{c}{B } \\ \hline
\hspace{0.5cm}&  Exact\hspace{0.7cm}& DMRG\hspace{0.7cm}& Exact\hspace{0.7cm}& 
 DMRG\hspace{0.7cm}&  Exact \hspace{0.7cm}& DMRG \hspace{0.7cm}&  Exact\hspace{0.7cm}& DMRG  \\ \hline 
1 \hspace{0.5cm}&  1.9083\hspace{0.7cm}&  1.9082\hspace{0.7cm}&  1.8795\hspace{0.7cm}&  1.8794\hspace{0.7cm}& 
   0.0096\hspace{0.7cm}&  0.0095\hspace{0.7cm}&  0.0059\hspace{0.7cm}&  0.0059 
\\
2 \hspace{0.5cm}&  1.7124\hspace{0.7cm}&   1.7123\hspace{0.7cm}&   1.7456\hspace{0.7cm}& 1.7455\hspace{0.7cm}&  
  0.0270\hspace{0.7cm}&  0.0269\hspace{0.7cm}&  0.0123\hspace{0.7cm}&  0.0123
 \\ 
3 \hspace{0.5cm}&  0.9449\hspace{0.7cm}&  0.9447\hspace{0.7cm}&   0.7752\hspace{0.7cm}& 0.7752\hspace{0.7cm}&  
  0.0745\hspace{0.7cm}&  0.0744\hspace{0.7cm}& 0.0570\hspace{0.7cm}&  0.0569 \\
4 \hspace{0.5cm}&  1.7491\hspace{0.7cm}&   1.7491\hspace{0.7cm}&   1.8383\hspace{0.7cm}&  1.8383\hspace{0.7cm}&
  -0.0045\hspace{0.7cm}&  -0.0045\hspace{0.7cm}&  -0.0025\hspace{0.7cm}&  -0.0025  \\
5 \hspace{0.5cm}&  1.1393\hspace{0.7cm}&   1.1392\hspace{0.7cm}&   1.0054\hspace{0.7cm}&  1.0054\hspace{0.7cm}& 
  0.0805\hspace{0.7cm}&  0.0806\hspace{0.7cm}&  0.1294\hspace{0.7cm}&  0.1295  \\
6 \hspace{0.5cm}&  1.7366\hspace{0.7cm}&   1.7365\hspace{0.7cm}&   1.8495\hspace{0.7cm}&  1.8495\hspace{0.7cm}& 
  0.0191\hspace{0.7cm}&  0.0191\hspace{0.7cm}&  0.0074\hspace{0.7cm}&  0.0074 \\
7 \hspace{0.5cm}&  1.1092\hspace{0.7cm}&  1.1091\hspace{0.7cm}&   1.0387\hspace{0.7cm}&  1.0387\hspace{0.7cm}& 
  -0.0297\hspace{0.7cm}&  -0.0299\hspace{0.7cm}&  -0.0854\hspace{0.7cm}&  -0.0856  \\
8 \hspace{0.5cm}&  1.7639\hspace{0.7cm}&   1.7639\hspace{0.7cm}&   1.8751\hspace{0.7cm}&  1.8752\hspace{0.7cm}& 
  0.0004\hspace{0.7cm}&  0.0004\hspace{0.7cm}&  0.0029\hspace{0.7cm}&  0.0029 \\
9 \hspace{0.5cm}& 1.1393\hspace{0.7cm}& 1.1394\hspace{0.7cm}& 1.0776\hspace{0.7cm}& 1.0777\hspace{0.7cm}& 
 0.1678\hspace{0.7cm}& 0.1680\hspace{0.7cm}& 0.2348\hspace{0.7cm}& 0.2351 \\
10 \hspace{0.5cm}& 1.7752\hspace{0.7cm}& 1.7753\hspace{0.7cm}& 1.8820\hspace{0.7cm}& 1.8819\hspace{0.7cm}& 
 0.0149\hspace{0.7cm}& 0.0150\hspace{0.7cm}& 0.0064\hspace{0.7cm}&  0.0064 \\
11 \hspace{0.5cm}& 1.1809\hspace{0.7cm}& 1.1812\hspace{0.7cm}& 1.1007\hspace{0.7cm}& 1.1010\hspace{0.7cm}& 
 -0.0771\hspace{0.7cm}& -0.0773\hspace{0.7cm}& -0.1229\hspace{0.7cm}& -0.1232 \\
12 \hspace{0.5cm}& 1.7508\hspace{0.7cm}& 1.7507\hspace{0.7cm}& 1.8763\hspace{0.7cm}& 1.8763\hspace{0.7cm}& 
 0.0081\hspace{0.7cm}& 0.0080\hspace{0.7cm}& 0.0049\hspace{0.7cm}& 0.0049 \\
13\hspace{0.5cm}& 1.1351 \hspace{0.7cm}& 1.1353 \hspace{0.7cm}& 1.0815 \hspace{0.7cm}& 1.0814 \hspace{0.7cm}& 
 0.1988\hspace{0.7cm}& 0.1991\hspace{0.7cm}& 0.2431\hspace{0.7cm}& 0.2433 \\
14\hspace{0.5cm}& 1.9550\hspace{0.7cm}& 1.9550\hspace{0.7cm}& 1.9745\hspace{0.7cm}& 1.9745\hspace{0.7cm}& 
 0.0108\hspace{0.7cm}& 0.0108\hspace{0.7cm}& 0.0068\hspace{0.7cm}& 0.0068\\
\hline
\end{tabular}
\end{center}

\vspace{0.6cm}
\pagebreak

Table 4. Comparison of DMRG bond orders for the 
 14 sites  $MX$ chain with exact calculation. 
$p$ refers to the (p, p+1)  bond and A and B correspond to parameter values
in table 2.
\vspace{0.3cm}
\begin{center}
\begin{tabular}{ccccc} \hline
$p$ \hspace{0.5cm}& \multicolumn{2}{c}{A}\hspace{0.8cm}& \multicolumn{2}{c}{B} \\ \hline
\hspace{0.5cm}&  Exact \hspace{0.8cm}& DMRG\hspace{0.8cm}&  Exact\hspace{0.8cm}& DMRG  \\  \hline 
 1 \hspace{0.5cm}& 0.1594\hspace{0.8cm}&  0.1596\hspace{0.8cm}&  0.1700\hspace{0.8cm}&     0.1702  \\
 2 \hspace{0.5cm}& 0.5438 \hspace{0.8cm}&  0.5440 \hspace{0.8cm}& 0.5468\hspace{0.8cm}&    0.5468 \\
 3 \hspace{0.5cm}& 0.4452\hspace{0.8cm}&  0.4452\hspace{0.8cm}& 0.3772\hspace{0.8cm}&    0.3772 \\
4 \hspace{0.5cm}&  0.3662\hspace{0.8cm}&  0.3660\hspace{0.8cm}&  0.2926\hspace{0.8cm}& 0.2926 \\
 5 \hspace{0.5cm}& 0.4068 \hspace{0.8cm}&  0.4068 \hspace{0.8cm}& 0.3198\hspace{0.8cm}& 0.3198   \\
\hline
\end{tabular}
\end{center}
\vspace{0.6cm}

Table 5.  Fractional stabilization of energy defined as 
$(E(\delta)-E(0))/E(0)$ of $MX$ chains
with respect to uniform chain for different parameter sets.
$I$: $U_M=1.5t$, $U_X=0.5t$, $\epsilon_M=0$, $\epsilon_X=-t $;
$II$:  $U_M=1.5t$, $U_X=0.5t$, $\epsilon_M=0$, $\epsilon_X=-2t$;
$III$: $U_M=2.5t$, $U_X=0.5t$, $\epsilon_M=0$, $\epsilon_X=-t$;
$IV$:  $U_M=2.5t$, $U_X=0.5t$, $\epsilon_M=0$, $\epsilon_X=-2t$ and
$V$:   $U_M=2.5t$, $U_X=0.5t$, $\epsilon_M=t$, $\epsilon_X=-2t$.
\vspace{0.3cm}
\begin{center}
\begin{tabular}{cccccc} \hline
$\delta$ \hspace{0.3cm}&  I \hspace{0.3cm} &        II\hspace{0.3cm}  &      III \hspace{0.3cm}  &      IV \hspace{0.3cm}&  V  \\ \hline
0.10 \hspace{0.3cm} & -0.0043\hspace{0.3cm}  &  -0.0127\hspace{0.3cm}   & -0.0007\hspace{0.3cm} &   -0.0017\hspace{0.3cm}  & -0.0560 \\
0.15 \hspace{0.3cm} & -0.0109 \hspace{0.3cm} &  -0.0319\hspace{0.3cm}  & -0.0026\hspace{0.3cm}  &  -0.0055\hspace{0.3cm}  & -0.0990 \\
0.20\hspace{0.3cm}  & -0.0210 \hspace{0.3cm}  & -0.0598\hspace{0.3cm} &  -0.0054\hspace{0.3cm}   & -0.0104\hspace{0.3cm} &  -0.1489 \\ \hline
\end{tabular}
\end{center}

\vspace{0.6cm}

Table 6. Energy (in units of $t$) for 
doped $MX$ chains of 35 units
with one and two holes as well as one and two electrons, for various
representative parameters of the Peierls-Hubbard model. $I$ corresponds 
to $U_M=1.5t$, $U_X=0.5t$ and $\epsilon_X=-t$ and $II$ to 
 $U_M=2.5t$, $U_X=0.5t$ and $\epsilon_X=-2t$.
\vspace{0.6cm}
\begin{center}
\begin{tabular}{ccccccccc} \hline
& \multicolumn{4}{c}{$I$}\hspace{0.5cm}& \multicolumn{4}{c}{$II$} \\ \hline
doping\hspace{0.5cm}& \multicolumn{2}{c}{$\epsilon_M=0.0$}\hspace{0.5cm}& \multicolumn{2}{c}{$\epsilon_M=t$ }\hspace{0.5cm}& \multicolumn{2}{c}{$\epsilon_M=0.0$} \hspace{0.5cm}& \multicolumn{2}{c}{$\epsilon_M=t$} \\ \hline
\hspace{0.5cm}& $\delta=0.0$ \hspace{0.5cm}& $\delta=0.1$ \hspace{0.5cm}& $\delta=0.0$ \hspace{0.5cm}& $\delta=0.1$ \hspace{0.5cm}& $\delta=0.0$
\hspace{0.5cm}& $\delta=0.1$ \hspace{0.5cm}& $\delta=0.0$ \hspace{0.5cm}& 
$\delta=0.1$ \\ \hline
2 holes\hspace{0.5cm}& -9.6308\hspace{0.5cm}&  -9.7157\hspace{0.5cm}& 
-9.3790\hspace{0.5cm}&  -9.1630\hspace{0.5cm}&  -12.2875\hspace{0.5cm}& -12.3646\hspace{0.5cm}& -13.5137\hspace{0.5cm}& -13.4442 \\
 1 hole \hspace{0.5cm}& -4.8361\hspace{0.5cm}&  -4.8944\hspace{0.5cm}& -4.6590\hspace{0.5cm}&  -4.5038\hspace{0.5cm}&  -6.1595\hspace{0.5cm}& -6.2060\hspace{0.5cm}& -6.8225\hspace{0.5cm}& -6.7591  \\ 
 1 electron\hspace{0.5cm}& 4.9375\hspace{0.5cm}&  4.9451\hspace{0.5cm}& 5.2900\hspace{0.5cm}& 5.3242\hspace{0.5cm}&  6.1935\hspace{0.5cm}& 6.2158\hspace{0.5cm}& 6.8850\hspace{0.5cm}& 6.9995 \\
 2 electrons\hspace{0.5cm}&10.3529\hspace{0.5cm}&  10.3245\hspace{0.5cm}&  10.7410\hspace{0.5cm}&  10.8901\hspace{0.5cm}& 13.9677\hspace{0.5cm}& 
13.9482 \hspace{0.5cm}& 13.9909 \hspace{0.5cm}&  14.0724 \\ \hline
\end{tabular}
\end{center}

\pagebreak
{\bf Figure Captions} \\
\begin{enumerate}
\item Plot of energy per $MX$ unit vs $1/N$ for different values of $\delta$
 for $U_M=1.5t$, $U_X=0.5t$, $\epsilon_{M}=0.0$ and $\epsilon_{X}=-t$.
(i) $\delta=0.0$ (square) (ii) $\delta=0.1$ (circle) (iii) $\delta=0.15$
(triangle) (iv) $\delta=0.2$ (diamond).

\item Charge density of $M$ and $X$ vs unit cell index. Open and filled
symbols are for $M$ and $X$ charge densities, respectively. Squares
represent the charge density for $U_M=1.5t$, $U_X=0.5t$, $\epsilon_M=0$, 
$\epsilon_{X}=-t$  and $\delta=0.1$ and cirlces for $U_M=2.5t$, $U_X=0.5t$,
$\epsilon_M=0$, $\epsilon_{X}=-2t$ and $\delta=0.1$.

\item Charge density of $M$ vs unit cell index for different values of 
 $\delta$,
 for  $U_M=1.5t$, $U_X=0.5t$, $\epsilon_M=0$ and $\epsilon_{X}=-t$.
(i) $\delta=0.0$ (square) (ii) $\delta=0.1$ (circle) (iii) $\delta=0.15$
(triangle) (iv) $\delta=0.2$ (diamond).

\item Charge density of $M$ vs unit cell index in the presence of site-diagonal
distortion. (i) $U_M=1.5t$, $U_X=0.5t$, $\epsilon_M=t$, $\epsilon_{X}=-t$ and
$\delta=0.1$ (square). (ii) $U_M=2.5t$, $U_X=0.5t$, $\epsilon_M=t$,
$\epsilon_{X}=-2t$ and $\delta=0.1$ (circle).

\item Bond order vs bond index. (i) $U_M=1.5t$, $U_X=0.5t$, $\epsilon_M=0$,
$\epsilon_{X}=-t$ and $\delta=0.1$ (square). 
(ii) $U_M=1.5t$, $U_X=0.5t$, $\epsilon_M=t$, $\epsilon_{X}=-t$ and 
$\delta=0.1$ (circle). 
(iii) $U_M=2.5t$, $U_X=0.5t$, $\epsilon_M=0$, $\epsilon_{X}=-2t$ and 
$\delta=0.1$ (triangle). 
(iv) $U_M=2.5t$, $U_X=0.5t$, $\epsilon_M=t$, $\epsilon_{X}=-2t$ and 
$\delta=0.1$ (diamond).

\item Spin density of $M$ vs unit cell index in different parameter
regions. (i) $U_M=1.5t$, $U_X=0.5t$, 
$\epsilon_M=0$, $\epsilon_{X}=-t$ and $\delta=0.0$ (square). 
(ii) $U_M=1.5t$, $U_X=0.5t$, $\epsilon_M=0$, $\epsilon_{X}=-t$ and 
$\delta=0.1$ (circle). 
(iii) $U_M=2.5t$, $U_X=0.5t$, $\epsilon_M=0$, $\epsilon_{X}=-2t$ and 
$\delta=0.1$ (triangle). 
(iv) $U_M=2.5t$, $U_X=0.5t$, $\epsilon_M=t$, $\epsilon_{X}=-2t$ and 
$\delta=0.1$ (diamond). Zeroth line is shown by dots.

\item Variation of (a) charge densities and (b) spin densities with 
site-diagonal
distortion, $\epsilon_M$, vs unit cell index for $U_M=2.5t$, $U_X=0.5t$,
$\epsilon_X=-2t$ and $\delta=0.1$. (i) $\epsilon_M=0.0$ (open square)
(ii) $\epsilon_M=1.0$ (open circle) (iii) $\epsilon_M=2.0$ (filled square)
(iv) $\epsilon_M=3.0$ (filled circle).

\item  Spin structure factor vs momentum, q (in degrees) for (a) 
$U_M=1.5t$, $U_X=0.5t$, $\epsilon_M=0$,
$\epsilon_{X}=-t$ and  $\delta=0.0$. 
(b) $U_M=1.5t$, $U_X=0.5t$, $\epsilon_M=0$, $\epsilon_{X}=-t$ and 
$\delta=0.2$. 
(c) $U_M=2.5t$, $U_X=0.5t$, $\epsilon_M=0$, $\epsilon_{X}=-2t$ and 
$\delta=0.2$. 
(d) $U_M=2.5t$, $U_X=0.5t$, $\epsilon_M=t$, $\epsilon_{X}=-2t$ and 
$\delta=0.1$.

\item Charge structure factor  vs q (in degrees) for (a) 
$U_M=1.5t$, $U_X=0.5t$, $\epsilon_M=0$,
$\epsilon_{X}=-t$ and  $\delta=0.0$. 
(b) $U_M=1.5t$, $U_X=0.5t$, $\epsilon_M=0$, $\epsilon_{X}=-t$ and 
$\delta=0.2$. 
(c) $U_M=2.5t$, $U_X=0.5t$, $\epsilon_M=0$, $\epsilon_{X}=-2t$ and 
$\delta=0.2$. 
(d) $U_M=1.5t$, $U_X=0.5t$, $\epsilon_M=t$, $\epsilon_{X}=-t$ and 
$\delta=0.1$.

\item 
Charge density of metal site vs unit cell index for a uniform $MX$
chain. (a) polaron and (b) bipolaron  for $U_M=1.5t$, $U_X=0.5t$, 
$\epsilon_M=0$ and $\epsilon_{X}=-t$. (c) and (d) for polaron and
bipolaron respectively, for $U_M=2.5t$, $U_X=0.5t$, $\epsilon_M=0$ and
$\epsilon_{X}=-2t$. (i) positively charged (square), (ii) neutral (circle) 
and (iii) negatively charged (triangle) systems,
in all the figures (a)-(d).

\item  Charge density and spin density of metal site for positively doped
$MX$ chain for  $U_M=1.5t$, $U_X=0.5t$, $\epsilon_M=0$ and
$\epsilon_{X}=-2t$. Charge density for (a)  polaron and (b) bipolaron.
Spin density for (c) polaron and (d) bipolaron. In all the figures (a)-(d),
(i) $\delta=0.0$ (square), (ii) $\delta=0.1$ (circle) and 
(iii) $\delta=0.2$ (triangle).

\item 
Charge density  of metal site for positively charged polaron (squares) and 
bipolaron (circles) for  $U_X=0.5t$ and $\delta=0.1$. Open symbols for 
$U_M=1.5t$ 
and $\epsilon_X=-t$ and filled symbols are for  $U_M=2.5t$ 
and $\epsilon_X=-2t$, (a) for $\epsilon_M=0.0$ and (b) for 
$\epsilon_M=t$.

\item  Difference in (a) charge density and (b) spin density  of metal sites 
with respect 
to neutral system for positively charged polaron and bipolaron for the same 
set of parameters as in fig.12 with the single $\epsilon_M=t$.

\item 
Charge and spin density  of metal sites with respect 
to neutral system for negatively charged (i) polaron and (ii) 
bipolaron in the presence of site-diagonal distortion. (a) and (c)
for $U_M=1.5t$, $U_X=0.5t$, $\epsilon_M=t$, $\epsilon_{X}=-t$ and 
$\delta=0.1$ for charge and spin density respectively. Similarly (b) and
(d) for $U_M=2.5t$, $U_X=0.5t$, $\epsilon_M=t$, $\epsilon_{X}=-2t$ and 
$\delta=0.1$ for charge and spin density respectively.

\item 
Bond order vs bond index for  bipolaron. (i) positively charged (squares), 
(ii) neutral (triangles)
and (iii) negatively charged (circles) 
 systems. For $U_M=1.5t$, $U_X=0.5t$,  $\epsilon_{X}=-t$ and 
$\delta=0.1$ and (a) for $\epsilon_M=0$ and (b) for $\epsilon_M=t$.
For $U_M=2.5t$, $U_X=0.5t$,  $\epsilon_{X}=-2t$ and 
$\delta=0.1$ and (c) for $\epsilon_M=0$ and (d) for $\epsilon_M=t$.
\end{enumerate}

\end{document}